\documentclass[reprint, superscriptaddress, amsmath,amssymb, aps, prresearch, longbibliography, floatfix]{revtex4-2}
\usepackage{amsmath}
\usepackage{graphicx}
\usepackage[normalem]{ulem}
\usepackage{bm}
\usepackage{natbib}
\usepackage{dsfont}
\usepackage{tikz}
\usepackage{epsfig}
\usepackage{feynmf}
\usepackage{blindtext, rotating}
\usepackage{mathtools}
\usepackage{dsfont}
\usepackage{subcaption}
\usepackage{physics}
\usepackage{amsfonts}
\usepackage{xcolor}
\usepackage{ragged2e}
\usepackage{siunitx}
\usepackage{comment}
\usepackage{soul}
\usepackage{empheq}
\usepackage{lipsum}
\usepackage{hyperref} 
\hypersetup{breaklinks=true, colorlinks=true, citecolor=blue, linkcolor=cyan, urlcolor=blue,filecolor=blue}

\usepackage[T1]{fontenc}

\usepackage{xcolor} 

\DeclareCaptionJustification{justified}{\justifying}

\DeclareSIUnit{\rad}{rad}

\definecolor{bright_blue}{HTML}{85C1E9}
\definecolor{middle_blue}{HTML}{2E86C1}
\definecolor{dark_blue}{HTML}{1B4F72}

\DeclareCaptionLabelFormat{rbrace}{#2)} 
\begin{document}

\title{Mesoscopic mechanical superpositions by gluing individual quantum systems}

\author{Thiago Guerreiro}
\email{thguerreiro@gmail.com}
\affiliation{Department of Physics, Pontifical Catholic University of Rio de Janeiro, Rio de Janeiro 22451-900, Brazil}

\begin{abstract}
We propose a protocol for preparing mechanical Schrödinger kittens -- mesoscopic quantum superpositions of coherent motional states of an optically levitated nanoparticle -- by adhering single electron time-bin states to its surface. Over short protocol timescales, coherence of the mesoscopic superposition survives the dominant decoherence mechanisms afflicting levitated systems, and can be observed by varying the phase of the time-bin electrons. Interference fringes can be detected with real-time, near-Heisenberg limited interferometry of photons scattered from the particle. This approach eliminates the need for coherent state expansion, dark potentials, and particle release-and-recapture mechanisms, providing a new route to test quantum mechanics in unprecedented scales.
\end{abstract}


\maketitle


\textit{Introduction.}--- Optical trapping has played an important role in our ability to detect quantum mechanical effects with single atoms and microscopic systems \cite{chu1998nobel}. Today, optical levitation of nano- and micron-sized objects holds great potential for testing quantum mechanics in the macroscopic domain \cite{bose2025massive}, largely due to recent developments in ground state cooling \cite{delic2020cooling, pontin2023simultaneous, magrini2021real, tebbenjohanns2021quantum, ranfagni2022two, piotrowski2023simultaneous, kamba2023nanoscale}, coherent wavepacket expansion \cite{rossi2025quantum, kamba2025quantum, muffato2024coherent, steiner2025free, tomassi2026accelerated}, and precision force metrology \cite{skrabulis2026nanomechanical, otabe2026time, tseng2026optomechanical}. A promising route to observing quantum effects with levitated nanoparticles is preparing non-Gaussian states of the center-of-mass (CoM) motion, exhibiting matter-wave interference. Generating these states requires subjecting the nanoparticle to a non-harmonic (non-linear) potential, which plays an analogous role to the slit in a double slit experiment \cite{clauser1997broglie}. However, due to the optical diffraction limit, spatial variations in optical potentials occur on the order of $\sim 1 \ \mu$m, orders of magnitude larger than the characteristic width of the particle's ground-state wavefunction, typically of $\sim10\text{ pm}$.  

Consequently, many proposals rely on coherent state expansion \cite{romero2011large, weiss2021large, neumeier2024fast}, followed by subsequent re-capture of the levitated object in a hybrid or optical trap \cite{bonvin2024hybrid, bonvin2024state, mattana2026trap}. Unfortunately, large spatial expansion severely accelerates decoherence. For instance, the off-diagonal elements of a density matrix subject to weak localization evolve according to $\langle x \vert \rho \vert x'\rangle \sim \exp(-\Lambda \vert x-x'\vert^{2} t)$ \cite{caves1987quantum}, making mesoscopic quantum experiments extremely challenging. 

\begin{figure}[ht!]
    \centering
    \includegraphics[scale=1.0]{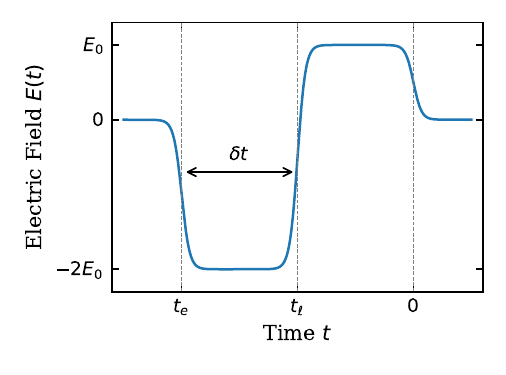}
    \caption{Time-dependent electric field for generating a coherent superposition of momentum kicks from the inelastic collision of a time-bin electron state with an initially neutral levitated particle. If the electron arrives at $t_{e}$, the particle receives a momentum kick of $\delta p = eE_{0}\delta t$, whereas if it arrives at $t_{\ell}$, the momentum kick is $-\delta p$. }
    \label{fig1}
\end{figure}

\begin{figure*}[ht!]
    \centering
    \includegraphics[scale=0.9]{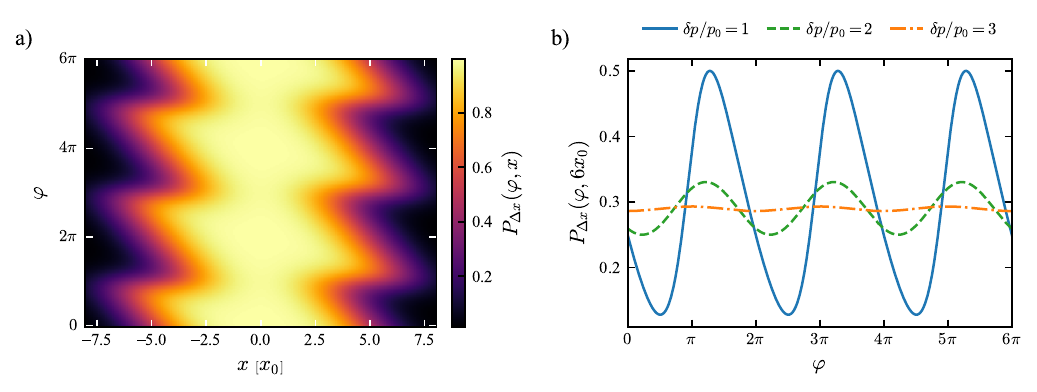}
    \caption{a) Interference in the position  probability  $P_{\Delta x}(\varphi,x)$ as a function of $\varphi$ for initial momentum kick $\delta p = p_{0}$, decoherence rate $ \Gamma / \omega = 0.2$, initial occupation number $\Bar{n} = 0.5$ and resolution $\Delta x = 10x_{0}$. b) Interference of position probability as a function of $\varphi$, at $x = 6x_{0}$ for different initial kicks $\delta p = p_{0}, 2p_{0}, 3p_{0}$, with the same remaining parameters as in a).}
    \label{interference-fringes}
\end{figure*}

Levitated particles are so well isolated from their environment \cite{dania2024ultrahigh} and so sensitive to external forces \cite{wang2024mechanical}, however, that one wonders whether the influence of an inherently quantum microscopic system upon the mesoscopic object could be used to effectively implement Schrödinger's cat gedanken experiment \cite{trimmer1980present}. One idea along those lines is to couple nanoparticles to atoms or ions  \cite{torovs2021creating, bykov2025nanoparticle, gupta2025quantum}, or to employ levitated particles with embedded quantum systems, such as NV centers in diamond \cite{yin2013large, conangla2018motion, levi2025quantum}. 

More direct -- as experimentally demonstrated \cite{Moore2014, frimmer2017controlling, stoellner2025using} -- is the capacity of a nanoparticle in an external field to respond to a single elementary charge. Noting that single electrons are inherently quantum mechanical and can be routinely prepared in coherent superposition states \cite{tonomura1989demonstration, bach2013controlled}, we explore a further possibility: combining optically trapped nanoparticles with coherent single-electron experiments to generate macroscopic quantum states of motion. Specifically, we propose a protocol that exploits the inelastic collision of a coherent single-electron wave-packet with an optically trapped nanoparticle, thereby producing a superposition of momentum kicks of the levitated nanoparticle, or a Schrödinger kitten state \cite{ourjoumtsev2006generating}. These states are non-classical, exhibiting Wigner negativity which survives for a considerable time during the protocol. By manipulating the phase of the electron's wavefunction, we can induce interference fringes in the nanoparticle's position probability distribution without directly acting on the particle. 
We analyze the various sources of decoherence and their impact on the interference, and show that high-visibilities can be achieved within a single oscillation period. The fringes can be detected by homodyne interferometry of photons scattered by the particle, thereby effectively realizing a mesoscopic matter-wave interference experiment of unprecedented scale. 

Remarkably, this approach is distinct from, and does not require coherent Bragg electron diffraction \cite{Nimmrichter2025ElectronEnabled}. It also bypasses the need for spatial state expansion, non-linear potentials, and particle release-and-recapture.

Throughout we consider a levitated particle in a harmonic trap with frequency $\omega$ along the longitudinal direction, aligned with the trapping beam wavevector. We denote the particle's zero-point position and momentum as $x_{0} = \sqrt{\hbar/(2m\omega)}, p_{0} = \sqrt{m\hbar \omega/2}$, respectively, and assume the harmonic motions in the longitudinal direction and transverse plane are decoupled. We take the particle to be an amorphous SiO$_{2}$ bead with a radius of $R = 50 $ nm in a trap with $\omega = 2\pi \times 100 $ kHz. The particle carries an initial charge given by $Q = Ze $, where $ e \approx -1.6\times 10^{-19}$~C is the elementary charge.

\textit{Protocol.}--- The protocol consists of the following steps:

\begin{itemize}
    \item[0)] Prepare the charged levitated particle in the ground state of the harmonic trap. 

    \item[1)] Prepare a single electron in a time-bin state
    \begin{eqnarray}
        \vert \psi_{e^{-}} \rangle = \frac{\vert t_{e}\rangle + e^{i\varphi}\vert t_{\ell}\rangle}{\sqrt{2}} 
        \label{time-bin}
    \end{eqnarray}
    where $\varphi$ is a controllable phase and $t_{e}, t_{\ell}$ denote early and late arrival times of the electron at the trapping site. We set the electron's energy such that once it collides with the particle's surface it gets trapped, decreasing its total charge by one elementary unit.

    \item[2)]  Apply a time-dependent electric field at the particle along the longitudinal direction, such as the one in Fig. \ref{fig1}, with $\delta t_{s} \ll \delta t \ll \frac{2\pi}{\omega} \ , \Gamma^{-1}$, where $\delta t_{s}$ is the electric field `switching time' and $\Gamma$ is the total decoherence rate of the system. The exact form of the profile depends on the initial charge $Z$ (see discussion below).

    \item[3)] Let the particle evolve in the harmonic potential until $ t = \pi / \omega$. 
    Then, measure the position of the particle $x$ with resolution given by $\Delta x$ over an integration time $\tau_{\rm meas} \ll 2\pi/\omega$. Repeat this step many times with a fixed value of $\varphi$ to reconstruct the probability $ P_{\Delta x}(\varphi,x)$ of finding the particle in the interval $ \left[ x-\frac{\Delta x}{2}, x + \frac{\Delta x}{2} \right] $.

    \item[4)] Repeat steps 0) - 3) varying the value of the relative phase $\varphi$. 
\end{itemize}


As an example, consider an initially neutral particle, $Z = 0$, and the profile shown in Fig. \ref{fig1}. We assume $\delta t$ is so small that the particle's harmonic motion and decoherence are approximatelly frozen during the state preparation, step 2). The applied electric field is such that if the electron collides with the particle at $t_{e} = -2\delta t$, it acquires a momentum kick of $\delta p_{e} = -\delta p$, while if it arrives at $t_{\ell} = -\delta t$, the momentum kick is $\delta p_{\ell} = +\delta p$, where $\delta p = \vert eE_{0} \vert \delta t $ and we consider $\delta t_{s} \approx 0$. Taking $\delta t = 10^{-2}\times (2\pi/\omega) = 100$ ns we have a momentum transfer of $\delta p = p_{0} $ for $E_{0} \approx 360$~V/m, easily achievable in experiments \cite{kremer2024all}. The electric field profile can be generalized to any initial particle charge $Q$, such that the particle always gets a kick of $\pm \delta p$, depending on whether the electron arrives at $t_{e}$ or $t_{\ell}$; see Appendix \ref{appendixA}.  

\begin{figure*}[ht!]
    \centering
    \includegraphics[scale=0.9]{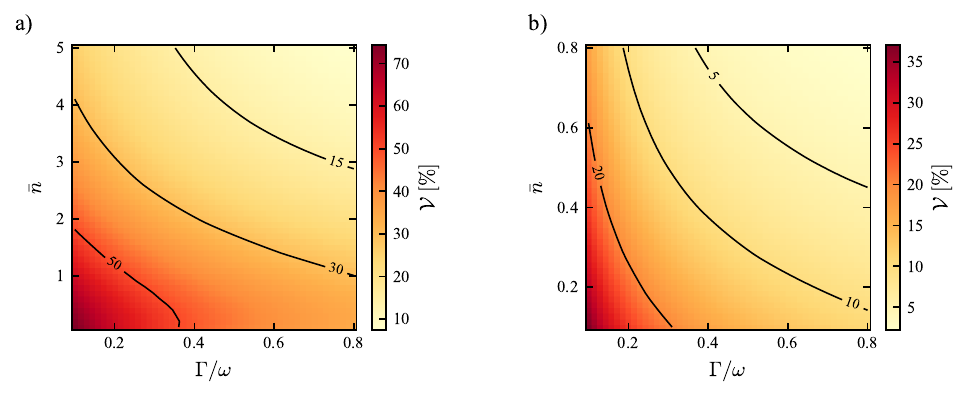}
    \caption{Interference visibility as a function of initial occupation number $\Bar{n}$ and decoherence rate $\Gamma$ for a) $\delta p = p_{0}$, b) $\delta p = 2p_{0}$. Remaining parameters are $ x = 6x_{0}$, $\Delta x = 10x_{0}, \eta = 1$.}
    \label{fig3}
\end{figure*}


Time-bin single-electron states can be prepared using electron microscopy technology \cite{tonomura1989demonstration}. For instance, coherent electronic superpositions are achievable via bi-prisms or the Kapitza-Dirac effect \cite{freimund2002bragg, lin2024ultrafast}, and electron Mach–Zehnder interference shows that high-purity time-bin states for energies between $50\text{ eV}$ and $1\text{ keV}$ can be produced \cite{barwick2006measurement, sonnentag2007measurement}. Recent advances further enable trapping and control of low-energy electrons \cite{matthiesen2021trapping}. For our protocol, the electron energy must be high enough to overcome the particle's Coulomb barrier and probe its dielectric surface image-charge potential, yet low enough to prevent decoherence by secondary electron emission. 

For a particle with charge $Ze$, we can roughly estimate the Coulomb barrier as the potential at the particle's radius \footnote{The attractive image charge potential will lower the critical barrier.},
\begin{eqnarray}
    U_{\rm c} \approx \frac{Ze^{2} }{4\pi\epsilon_{0} R} \approx (1.44 \ \text{eV}) \times Z \times  \left(\frac{1 \text{nm}}{R}\right) \ .
\end{eqnarray}
As an example, for $Z = 250 $ we have $U_{c} \approx 7 \ \text{eV} $, so we need the time-bin electron states to have energy $U_{\rm time-bin} \gtrsim 7 \ \text{eV} $. Note the secondary electron emission energy threshold for amorphous SiO$_{2}$ is $ \approx 10 \ \text{eV} $ \cite{astavsauskas2020optical}, so we require 7 eV $ \lesssim U_{\rm time-bin} \lesssim 17 $ eV. At the end of each round of the protocol, we can check if the particle's charge decreased by one elementary unit using standard techniques \cite{Tandeitnik2026Heterodyne}, which heralds a successful measurement. 

Collision with the electron and application of the field $E(t)$ prepares -- in the ideal case -- the non-Gaussian state at $t=0$,
    \begin{eqnarray}
        \vert \psi_{0}\rangle = N \left(e^{i\frac{\delta p \hat{x}}{\hbar}} + e^{i\varphi}e^{-i\frac{\delta p \hat{x}}{\hbar}}\right) \vert 0 \rangle
        \label{pure-state}
    \end{eqnarray}
where $N = \left(2 + 2\cos\varphi e^{-\delta p^{2}/2p_{0}^{2}}  \right)^{-1/2}$ is the state's normalization. Observe that for $\varphi = \pi$ this state has a negative Wigner function, with $W(x=0,p=0) = -2/\pi$ irrespective of the value of $\delta p$; see below for a discussion of how this negativity evolves under decoherence. For small values of $\delta p$, it also approximates the Fock state $\vert 1 \rangle$, and can exhibit superoscillations \cite{aharonov1990superpositions}. During steps 2) and 3), $\vert \psi_{0}\rangle$ evolves to a superposition of Gaussians centered at distinct locations, reaching a maximum separation of $\delta x = \frac{2\delta p}{m\omega}$ at $t=\pi/2\omega$ and overlapping again at $ t = \pi/\omega$. 

The result from steps 3) - 4) is an interference pattern in the particle's CoM position probability as a function of $\varphi$. At half a period of oscillation, $ t = \pi/\omega$, we have
\begin{eqnarray}
    P_{\Delta x}( \varphi,x) = N^{2}\left(A + B\cos\varphi + C\sin \varphi \right)
    \label{pure-prob}
\end{eqnarray}
where $A, B$ and $C$ are coefficients that depend on the position and resolution (see below). Varying the phase $\varphi$ of $\vert \psi_{e^{-}} \rangle$ at successive runs of the protocol reveals fringes in the particle's CoM probability distribution. Note that placing control over the phase $ \varphi$ at the electron relaxes the requirement of actuating directly on the particle's state, which is more susceptible to decoherence due to its macroscopic size.

\textit{Decoherence.}--- The coefficients in eq. \eqref{pure-prob} depend on the interaction of the particle with its environment, which causes loss of interference contrast. We now consider various sources of decoherence and evaluate their impact on the interference visibility. 

Under UHV conditions (at pressures around $10^{-10}$ mbar) the probability of collision with residual gas molecules during a time interval of $ \Delta t = \pi/\omega $ is $ p_{\rm gas}< 10^{-3}$ for particles with a radius of $R=$ 50~nm in a trap with $\omega = 2\pi \times 100$~kHz. Therefore, gas collisions decoherence is negligible \cite{neumeier2024fast}.  

Another potential source of decoherence are the particle's internal modes of vibration: if the particle undergoes significant deformation upon collision with the electron, internal phonons might act as an environment capable of detecting ``which-path'' information \cite{henkel2023limit}. To evaluate that, we estimate the magnitude of deformations upon impact. Assuming a harmonic oscillator model for internal modes of vibration, the energy associated to a deformation of characteristic length $\delta R$ is given by $  E_{s} = m_{s}\omega_{s}^{2}\delta R^{2}/2$, where $ m_{s}$ and $\omega_{s} $ are the internal mode's effective mass and frequency. Equating this energy to the electron's kinetic energy, $E_{s} \approx \Delta p^{2}/(2m_{e})$, we may estimate $\delta R$. Assuming $ m_{s} = m $, $\omega_{s} = 10 \ \text{GHz} $ \cite{kuok2003brillouin} and $\Delta p = \sqrt{2m_{e} \times (10 \ \text{eV})} $ yields $\delta R \approx a_{0} $, where $ a_{0} $ is Bohr's radius. Hence, deformations due to the impact of the electron are on the same order as the uncertainty of the position of electrons in the particle's atoms, and therefore negligible.

Next, we have scattering of photons from the trapping laser \cite{jain2016direct}, the emission of black-body radiation from the particle \cite{hackermuller2004decoherence} and trap position and frequency fluctuations \cite{weiss2021large,aspelmeyer2022zeh}. All of these decoherence mechanisms are described by the master eq.,
\begin{eqnarray}
    \dot{\rho} = -\frac{i}{\hbar}[H,\rho] - \frac{\Lambda}{2} [x,[,x,\rho]]
    \label{master-eq}
\end{eqnarray}
where $\Lambda$ is the localization rate, related to the heating rate according to $ \Gamma = \Lambda x_{0}^{2} $. The decoherence rate for thermal scattering, emission and absorption of photons can be estimated to be $\Gamma_{\text{bb}} \lesssim 1 \ \text{Hz}$, much smaller than $\omega$. Similarly, trap position and frequency fluctuations yield a decoherence rate of $\Gamma_{\text{fluct}} \lesssim 1 \ \text{kHz}$, also significantly smaller than $\omega$ (see Appendix \ref{appendix-decoherence} for details). The main source of decoherence is then recoil heating from the trapping laser beam, which for a $R = 50$ nm particle yields a rate $ \Gamma_{\text{rc}} \approx 2\pi \times 20 \ \text{kHz} $ \cite{jain2016direct, magrini2021real}. From now on we take the total decoherence rate to be approximatelly $ \Gamma \approx \Gamma_{\text{rc}} $, or $ \Gamma / \omega \approx 0.2 $.

Assuming the initial state in step 0) is a thermal state with occupation number $\Bar{n}$, we can solve \eqref{master-eq} and calculate the $ABC$ coefficients defining $P_{\Delta x}(\varphi,x)$ in \eqref{pure-prob}. These are,
\begin{eqnarray}
    A &=& \erf\left( \frac{ x+\Delta x/2}{\sqrt{2}x_{0}\mu} \right) - \erf\left( \frac{x-\Delta x/2}{\sqrt{2}x_{0}\mu} \right) \label{A} \\
    B &=&  \eta  \frac{e^{-\mathcal{D}}}{\mu} \sqrt{\frac{2}{\pi}}\int_{\mathcal{I}} dy \cos\left(\frac{\xi^{2}}{\mu^{2}}\frac{\delta p}{p_{0}}y\right) e^{-y^{2}/2\mu^{2}}
    \label{B} \\
    C &=&  -\eta  \frac{e^{-\mathcal{D}}}{\mu} \sqrt{\frac{2}{\pi}}\int_{\mathcal{I}} dy \sin\left(\frac{\xi^{2}}{\mu^{2}}\frac{\delta p}{p_{0}}y\right) e^{-y^{2}/2\mu^{2}}
    \label{C} 
\end{eqnarray}
where $\xi = \sqrt{2\Bar{n}+1}$, 
\begin{eqnarray}
    \mu &=& \sqrt{\xi^{2} + \frac{2\pi \Gamma}{\omega}} \ , \\
    \mathcal{D} &=& \left( 1 - \frac{\xi^{2}}{\mu^{2}}  \right)  \frac{\xi^{2}\delta p^{2}}{2p_{0}^{2}} \ ,
\end{eqnarray}
the integration interval is $ \mathcal{I} = \left[ \frac{x - \Delta x/2}{x_{0}} , \frac{x + \Delta x/2}{x_{0}} \right]$, and $\eta$ is a parameter which accounts for dephasing in the electron time-bin state. The normalization factor reads $ N = \left(2 + 2\eta \cos\varphi e^{-\xi^{2}\delta p^{2}/2p_{0}^{2}}  \right)^{-1/2} $. 
We refer to the Appendix \ref{appendix-master} for details.

To understand the effects of recoil heating decoherence upon the levitated nanoparticle, we first assume a perfect electron time-bin state ($\eta = 1$), but see below for a discussion on the effects of dephasing on the electron wavefunction. Fig. \ref{interference-fringes}a) shows the probability distribution $P_{\Delta x}(\varphi,x)$ (at $t = \pi/\omega$) displaying clear interference fringes for a measurement resolution of $\Delta x = 10x_{0} \approx 85 \ \text{pm}$, an initial momentum kick $\delta p = p_{0}$, occupation number $\Bar{n} = 0.5$ and decoherence rate $\Gamma = 0.2\omega$. This probability displays an asymmetry in position characteristic of twisted cat states \cite{bild2023schrodinger}. The main effect of decoherence is to decrease the contrast of the interference. Importantly, the larger the initial momentum kick -- i.e. the more distinguishable the initial state -- the more severe is the reduction in contrast, as can be seen in Fig. \ref{interference-fringes}b). 

\begin{figure}[t!]
    \centering
    \includegraphics[scale=0.9]{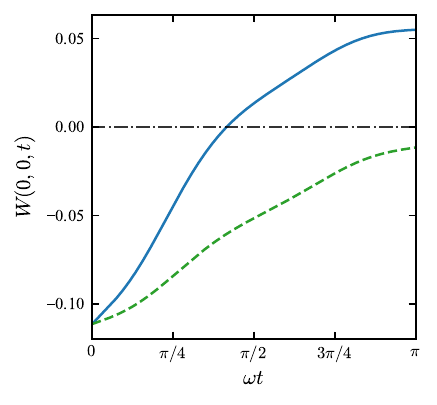}
    \caption{Time evolution of the Wigner function at the origin of phase space $W(x=0,p=0,t)$ for an initial state with $\bar{n} = 0.5$, $\delta p= p_{0}$, $\eta = 1$, and $\varphi = \pi$, displaying negative values, i.e. non-classicality over the course of the protocol. We consider two values for the decoherence: $\Gamma/\omega = 0.2$ (solid blue line), and $\Gamma/\omega = (1/3)\times 0.2$ (dashed green line), with the latter recoil heating rate enabled by a radially polarized optical trap.}
    \label{fig4}
\end{figure}

Fig. \ref{fig3} shows plots of the visibility , $\mathcal{V}~=~(\mathrm{max} \ P_{\Delta x} - \mathrm{min} \ P_{\Delta x})/(\mathrm{max} \ P_{\Delta x} + \mathrm{min} \ P_{\Delta x})$, as a function of initial occupation number $\bar{n}$ and total decoherence rate $\Gamma$, for momentum kicks given by $\delta p = p_{0}$ and $\delta p = 2p_{0}$. For $ \Gamma/\omega \approx 0.2 $ and $\bar{n} \approx 1$ we can expect visibilities close to 50\% for $\delta p = p_{0} $ and 10\% for $\delta p = 2p_{0} $. We also note the interference contrast is very robust against imperfections in the preparation of the initial ground state, as well as against decoherence. This robustness comes from the small size of the particle's wavepacket and from the fast protocol time of $\pi/\omega$. 


\textit{Time-bin dephasing.}--- We next take dephasing of the electron's wavefunction into account by replacing the time-bin state $\vert \psi_{e^{-}}\rangle$ by a density matrix with off-diagonal element $\langle t_{e} \vert \rho_{e^{-}} \vert t_{\ell}\rangle = \eta e^{i\varphi}$, where $\eta \in \left[ 0,1\right]$. The parameter $\eta$ can be directly interpreted as the visibility in an electron Mach-Zehnder interferometer. As it turns out, for the same parameters as considered in Fig. \ref{fig3}a) the visibility scales linearly with $\eta$, as $\mathcal{V} \approx (50\% ) \times \eta$. An electron interference experiment with visibility of $\eta = 20\%$ would then yield a nanoparticle interference visibility of $\mathcal{V} \approx 10\%$. 


\textit{Wigner negativity.}--- For $\varphi = \pi$, the state \eqref{pure-state} exhibits Wigner negativity -- a signature of non-classicality. Although reduced by the finite occupation number of the initial state and decoherence, this negativity persists for a period during the protocol, as can be seen in Fig. \ref{fig4}, showing the value of the Wigner function at the phase-space origin as a function of time, $W(0,0,t)$. For $\Gamma/\omega = 0.2$, negativity survives until $\omega t \approx 5\pi / 12$. Reducing decoherence threefold -- e.g., using radially polarized vector beams \cite{almeida2025levitated, tandeitnik_position_prep} -- preserves negativity throughout the entire protocol. 


\textit{Detection.}--- Finally, we point out that optical homodyne detection of scattered photons from the trapping laser enables real-time detection of the particle's position near the Heisenberg limit \cite{magrini2021real}. The measurement resolution is given by $\Delta x \approx \sqrt{S_{\rm imp}/(2\tau_{\rm meas})}$, where $S_{\rm imp}$ is the total imprecision noise power spectrum and $\tau_{\rm meas}$ is the measurement integration time \cite{jain2016direct, tebbenjohanns2019optimal}. For the imprecision of $\sqrt{S_{\rm imp}} \approx 10^{-14} \ \text{m}/\sqrt{\text{Hz}}$ reported in Ref. \cite{magrini2021real}, a measurement resolution of $\Delta x \sim x_{0}$ can be achieved in $\tau_{\rm meas} \approx 2.7 \ \mu$s. For $\Delta x \sim 10x_{0}$, as considered above, this integration time reduces to $\tau_{\rm meas} \approx 27$ ns -- much shorter than the particle's 10 $\mu$s oscillation period. Continuous interferometric measurements of the particle's CoM position will directly reveal the interference fringes.



\textit{Conclusion.}--- In conclusion, we have proposed a protocol to prepare and detect mesoscopic center-of-mass quantum superpositions of a levitated nanoparticle. By binding coherent electron time-bin states onto the particle's surface, external electric fields can impart superpositions of momentum kicks. Controlling the relative phase of the electron time-bin wavefunction generates interference fringes in the particle's position probability distribution without directly acting on the mesoscopic system and without the need for coherent state expansion. This interference can subsequently be read out via real-time optical interferometry of photons scattered from the particle. Notably, the ability of a microscopic quantum object (the electron) to influence a mesoscopic particle many orders of magnitude more massive ($m/m_{\text{e}} \sim 10^{12}$) is a direct manifestation of the strong character of the electromagnetic force. If realized, this approach offers a path toward quantum superpositions with objects at least five times more massive than the current state of the art~\cite{pedalino2026probing}, enabling tests of quantum mechanics in previously unexplored regimes \cite{penrose2014gravitization}.



\acknowledgments{
We acknowledge Antonio Zelaquett Khoury, Joanna A. Zielińska, Hendrik Ulbricht, George Svetlichny, Daniel Tandeitnik and Bruno Melo for conversations. 
We acknowledge support from the Coordenac\~ao de Aperfei\c{c}oamento de Pessoal de N\'ivel Superior - Brasil (CAPES) - Finance Code 001, the Brazilian National Institute of Science and Technology in Quantum Devices (INCT-DQ) and the Brazilian National Council for Scientific and Technological Development (CNPq, Grant No. 408783/2024-9), Funda\c{c}\~ao de Amparo \`a Pesquisa do Estado do Rio de Janeiro (FAPERJ Scholarships No. E-26/200.251/2023, E-26/210.249/2024, E-26/210.824/2025 and E-26/210.373/2026), Funda\c{c}\~ao de Amparo \`a Pesquisa do Estado de São Paulo (FAPESP processo 2021/06736-5), the Serrapilheira Institute (grant
No. Serra – 2211-42299) and StoneLab.}

\bibliography{main}

\appendix

\onecolumngrid

\section{Electric actuation for arbitrarily charged particles}\label{appendixA}

\begin{figure}[h]
    \centering
    \includegraphics[scale=1.0]{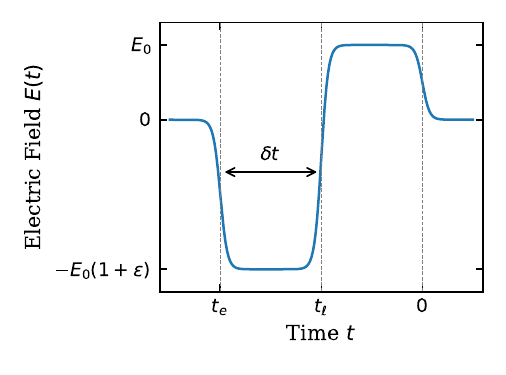}
    \caption{Electric field profile for producing a superposition of positive and negative momentum kicks $\pm \delta p$ for an arbitrary initial charge $Z$ of the nanoparticle. }
    \label{fig1_appendix}
\end{figure}

Consider that the harmonic oscillator has an initial charge given by $Q = Z e$, and we apply the electric field shown in Fig.  \ref{fig1_appendix}. If the electron hits the particle at time $t_{e}$, the particle acquires a momentum kick 
\begin{eqnarray}
    \delta p_{e} = -(Z+1)eE_{0}(1+\epsilon)\delta t + (Z+1) e E_{0}\delta t = -eE_{0}\delta t \epsilon (1+Z)
\end{eqnarray}
On the other hand, if the electron arrives at $t_{\ell}$, the momentum kick is
\begin{eqnarray}
    \delta p_{\ell} = -ZeE_{0}(1+\epsilon) \delta t + (Z+1)eE_{0}\delta t = - eE_{0}\delta t (1 - Z\epsilon)
\end{eqnarray}
We want to have $\delta p_{e} = - \delta p_{\ell}$, which is satisfied if 
\begin{eqnarray}
    \epsilon = \frac{1}{1+2Z}
\end{eqnarray}
In this case, we have 
\begin{eqnarray}
     \delta p_{\ell} = -\delta p_{e}  =   eE_0\delta t \left(\frac{1+Z}{1+2Z}\right)
\end{eqnarray}
Note that as $ Z \rightarrow \infty$, the momentum kick goes to $\vert \delta p \vert \rightarrow eE_{0}\delta t / 2$; considering the same parameters as in the main text, we achieve 
$\vert \delta p \vert \sim p_{0}$ for $E_{0}\sim 720$ V/m.

\section{Black-body and trap stability decoherence}\label{appendix-decoherence}

\textit{Thermal emission decoherence.} As discussed in the main text, the particle emits and absorbs black-body photons, which leads to decoherence. The typical wavelength of thermal photons is given by \cite{romero2011quantum}
\begin{eqnarray}
    \lambda_{\text{bb}} =\frac{\pi^{2/3}\hbar c}{k_B T}
\end{eqnarray}
which for temperatures in the range $T\sim$ 300 K to 1000 K is between $\lambda_{bb}\sim$ 16 $\mu$m to 5 $\mu$m. Since the wavelength of thermal photons is much larger than the particle's radius $\lambda_{bb} \gg R = 50$ nm, decoherence due to thermal radiation is well described by the master eq. \eqref{master-eq} in the main text. The three contributions to thermal decoherence are scattering, emission and absorption of thermal photons, with localization rates given by
\begin{equation}
\Lambda_{\text{bb}}^{sc} = \frac{8! \times 8\zeta(9)cR^6}{9\pi} \left[ \frac{k_B T_e}{\hbar c} \right]^9 \text{Re} \left[ \frac{\epsilon_{\text{bb}} - 1}{\epsilon_{\text{bb}} + 2} \right]^2
\end{equation}
for scattering and
\begin{equation}
\Lambda_{\text{bb}}^{e(a)} = \frac{16\pi^5 c R^3}{189} \left[ \frac{k_B T_{i(e)}}{\hbar c} \right]^6 \text{Im} \left[ \frac{\epsilon_{\text{bb}} - 1}{\epsilon_{\text{bb}} + 2} \right] .
\end{equation}
for emission and absorption. Here $\epsilon_{\text{bb}}$ is the dielectric constant, $\zeta$ is the zeta function and $T_{i}, T_{e}$ are the particle's internal temperature and environmental temperature, respectively. Considering the worst-case scenario of $ \text{Re} [ (\epsilon_{\text{bb}} - 1)/(\epsilon_{\text{bb}} + 2)] \sim \text{Im} [(\epsilon_{\text{bb}} - 1)/(\epsilon_{\text{bb}} + 2)] \approx 1 $ \cite{aspelmeyer2022zeh}, $R= 50$ nm, $ T_{i} = 1000$ K \cite{hebestreit2018measuring}, and $T_{e} = 300 $ K, we find
\begin{eqnarray}
    \Lambda_{\text{bb}}^{sc} &\sim & 6 \times 10^{14} \ \text{Hz/m}^{2} \\
    \Lambda_{\text{bb}}^{e} 
    &\sim  & 7 \times 10^{21} \ \text{Hz/m}^{2} 
\end{eqnarray}
from which we see thermal decoherence is dominated by emission of photons from the particle. We thus have a thermal decoherence rate of 
\begin{eqnarray}
    \Gamma_{\text bb} \sim \Lambda_{\text{bb}}^{e} x_{0}^{2} \sim 0.5 \ \text{Hz} \ ,
\end{eqnarray}

\textit{Trap fluctuations decoherence.} Another source of decoherence are fluctuations in trap position and frequency. These can be estimated as \cite{weiss2021large, aspelmeyer2022zeh},
\begin{eqnarray}
    \Gamma_{x} &=& \pi \omega^{2} S_{x}(\omega) / (4x_{0}^{2}) \\
    \Gamma_{\omega} &=& \pi \omega^{2} S_{\omega}(2\omega) / 16
\end{eqnarray}
where $S_{x,\omega}$ are the power spectral densities of position and frequency fluctuations, bounded by experiments according to $\sqrt{X_{x}(\omega)} < 10^{-16} \ \text{m}/\sqrt{\text{Hz}}$ and $ S_{\omega}(2\omega) < 10^{-4} \ /\sqrt{\text{Hz}}$. Therefore, we have the value used in the main text $\Gamma_{\text{fluct}} \approx \Gamma_{\omega} + \Gamma_{x} \lesssim 1 \ \text{kHz} $.

\section{Master equation}  \label{appendix-master} 

The eq. of motion for our system reads,
\begin{equation}
    \dot{\rho} = -\frac{i}{\hbar}[H,\rho] + \frac{\Lambda}{2}[x,[x,\rho]]
    \label{lindblad-appendix}
\end{equation}
where $\Lambda$ is the localization rate, which has dimensions of $ L^{-2}T^{-1}$, and $H = \frac{p^{2}}{2m} + \frac{m\omega^{2}x^{2}}{2}$ is the harmonic oscillator's Hamiltonian. From \eqref{lindblad-appendix} we can derive the eq. of motion for the phonon number operator $N = a^{\dagger}a$, given that jump operator is proportional to the position operator $x = x_{0}(a+a^{\dagger})$. We find 
\begin{eqnarray}
    \dot{N} = \Lambda x_{0}^{2} = \Gamma
\end{eqnarray}
where $\Gamma$ is the phonon heating rate due to particle localization, e.g. recoil heating \cite{jain2016direct}, as defined in the main text. 

The Wigner function reads 
\begin{eqnarray}
    W(x,p) = \frac{1}{\pi \hbar} \int dy \rho(x-y,x+y)e^{2ipy/\hbar}
\end{eqnarray}
where $\rho(x-y,x+y) = \langle x-y\vert \rho \vert x+y\rangle$ and integrals run from $-\infty$ to $+\infty$, unless specified. Eq. \eqref{lindblad-appendix} implies the eq. of motion for the Wigner function
\begin{equation}
    \frac{\partial W}{\partial t} = -\frac{p}{m} \frac{\partial W}{\partial x} + m\omega^{2}x\frac{\partial W}{\partial p} + \frac{\Lambda \hbar^{2}}{2} \frac{\partial^{2}W}{\partial p^{2}}
    \label{Wigner-dynamics-appendix}
\end{equation}

Now, define the Fourier transform of the Wigner function, or characteristic function
\begin{equation}
    \chi(u,v) = \int \int dx dp W(x,p) e^{-i(ux+vp)/ \hbar} = \int dx \rho(x-\frac{v}{2}, x+ \frac{v}{2}) e^{-iux/\hbar}
    \label{characteristic-FT}
\end{equation}
Substituting this expression in eq. \eqref{Wigner-dynamics-appendix} and performing integration by parts neglecting boundary terms we find,
\begin{equation}
    \frac{\partial \chi}{\partial t} + m\omega^{2}v \frac{\partial \chi}{\partial u} - \frac{u}{m}\frac{\partial \chi}{\partial v} = - \frac{\Lambda}{2} v^{2} \chi
    \label{main-eq-motion}
\end{equation}
which can be solved by the method of characteristics \cite{arnold1989mathematical}. From the solution of this eq., we can obtain the position probability density of the particle's CoM by inverse Fourier transforming the characteristic function,
\begin{equation}
    P(x,t) = \frac{1}{2\pi \hbar} \int du \chi(u,0,t) e^{iux/\hbar}
    \label{position-prob}
\end{equation}
Inverting \eqref{characteristic-FT} we may write the Wigner function in terms of the characteristic function, from which we obtain the value of the Wigner function at the origin in phase space,
\begin{eqnarray}
    W(0,0,t) =\frac{1}{(2\pi \hbar)^{2}}\int \int du dv \chi(u,v,t)
    \label{Wigner-negativity}
\end{eqnarray}
Let us solve \eqref{main-eq-motion} and compute \eqref{position-prob} and \eqref{Wigner-negativity}.

From the method of characteristics,
\begin{equation}
    \frac{d\chi}{dt} = \frac{\partial \chi}{\partial t} + \frac{du}{dt}\frac{\partial \chi}{\partial u}+ \frac{dv}{dt}\frac{\partial \chi}{\partial v} = - \frac{\Lambda}{2} v^{2} \chi
\end{equation}
where 
\begin{eqnarray}
    \frac{du}{dt} &=& m\omega^{2}v \label{characteristic1}\\ \frac{dv}{dt} &=& -\frac{u}{m} \label{characteristic2} \\
    \frac{d\chi}{dt} &=& - \frac{\Lambda}{2} v^{2} \chi \label{characteristic3}
\end{eqnarray}
Solving \eqref{characteristic1} and \eqref{characteristic2}, 
\begin{equation}
    u(t) = u_{0}\cos\omega t - m\omega v_{0}\sin\omega t \ , \ v(t) = -\frac{u_{0}}{m\omega}\sin \omega t + v_{0} \cos \omega t \label{characteristic-solutions}
\end{equation}
we can invert these eqs. and write $u_{0}, v_{0}$ in terms of $u,v$,
\begin{eqnarray}
    u_{0}(u,t) &=& u\cos \omega t  - m\omega v \sin \omega t \\
    v_{0}(u,v) &=& \frac{u}{m\omega}\sin\omega t  + v\cos \omega t 
\end{eqnarray}
Putting it all together,
\begin{equation}
    \chi(u,v,t) = \chi_{0}(u_{0}(u,v,t),v_{0}(u,v,t)) e^{-\frac{\Lambda}{2}\int_{0}^{t}dt' v^{2}(t')}
    \label{characteristic-function-solution}
\end{equation}
where $\chi_{0}(u_{0},v_{0})$ is the characteristic function of the initial state and
\begin{eqnarray}
    \exp\left({-\frac{\Lambda}{2}\int_{0}^{t}dt' v^{2}(t')}\right) = \exp\left(  -\alpha(\omega t) \left(\frac{u}{p_{0}} \right)^{2}  -\beta(\omega t) \left(\frac{v}{x_{0}} \right)^{2}  -\gamma(\omega t) \left(\frac{uv}{x_{0}p_{0}} \right) +   \right) 
\end{eqnarray}
with
\begin{eqnarray}
    \alpha(\tau) &=& \frac{\Gamma}{4\omega} \left( \tau - \frac{1}{2}\sin 2\tau \right) \\
    \beta(\tau) &=& \frac{\Gamma}{4\omega} \left( \tau + \frac{1}{2}\sin 2\tau \right) \\
    \gamma(\tau) &=&  \frac{\Gamma}{4\omega} \sin^{2} \tau
\end{eqnarray}

The density matrix for a thermal state with occupation number $\Bar{n}$ reads
\begin{eqnarray}
    \rho_{th}(x,y) = \frac{1}{\sqrt{2\pi x_{0}^{2}\xi^{2}}} \exp\left( -\frac{(x+y)^{2}}{8x_{0}^{2}\xi^{2}} - \frac{\xi^{2}(x-y)^{2}}{8x_{0}^{2}}  \right)
    \label{thermal}
\end{eqnarray}
where $\xi = \sqrt{2\Bar{n}+1}$. We consider that at $t=0$ the particle is prepared in the state 
\begin{eqnarray}
    \rho_{0} = N^{2} \left( e^{i\frac{\delta p \hat{x}}{\hbar} } + e^{i\varphi} e^{-i\frac{\delta p \hat{x}}{\hbar} }\right) \rho_{th} \left( e^{-i\frac{\delta p \hat{x}}{\hbar} } + e^{-i\varphi} e^{i\frac{\delta p \hat{x}}{\hbar} }\right)
    \label{initial_state_appendix}
\end{eqnarray}
where $N$ is the normalization factor defined in the main text. Combining this expression with \eqref{thermal} we obtain the initial state's components
\begin{eqnarray}
    \rho_{0}(x+\frac{v_{0}}{2}, x - \frac{v_{0}}{2}) = \frac{2N^{2}}{\sqrt{2\pi x_{0}^{2}\xi^{2}}} \left[  \cos\left(  \frac{\delta p v_{0}}{\hbar} \right) +  \cos\left(  \frac{2\delta p x}{\hbar} - \varphi \right) \right] e^{-\frac{x^{2}}{2x_{0}^{2}\xi^{2}} - \frac{\xi^{2}v_{0}^{2}}{8x_{0}^{2}}}
\end{eqnarray}
Fourier transforming \eqref{characteristic-FT} we find,
\begin{eqnarray}
    \chi_{0}(u_{0},v_{0}) = 2N^{2} \cos\left(  \frac{\delta p v_{0}}{\hbar} \right) e^{-\frac{\xi^{2}u_{0}^{2}}{8p_{0}^{2}} - \frac{\xi^{2}v_{0}^{2}}{8x_{0}^{2}}} + N^{2} e^{-\frac{\xi^{2}v_{0}^{2}}{8x_{0}^{2}}} \left(  e^{-i\varphi} e^{-\frac{\xi^{2}(u_{0}-2\delta p)^{2}}{8p_{0}^{2}}} + e^{i\varphi} e^{-\frac{\xi^{2}(u_{0}+2\delta p)^{2}}{8p_{0}^{2}}}  \right)
    \label{initial-characteristic-function}
\end{eqnarray}
Directly substituting in \eqref{characteristic-function-solution} and computing the Fourier transform \eqref{position-prob} we obtain
\begin{eqnarray}
    P(x,t=\pi/\omega) = \frac{N^{2}}{x_{0}\mu}\sqrt{\frac{2}{\pi}} e^{-x^{2}/(2x_{0}^{2}\mu^{2})}\left(  1 + e^{-\mathcal{D}}\cos\left( \varphi + \frac{2\xi^{2}}{\mu^{2}}\frac{\delta p x}{\hbar}\right)  \right)
\end{eqnarray}
where $\mu, \mathcal{D}$ are defined in the main text. The probability function used throughout the main text can then be obtained by integrating
\begin{eqnarray}
    P_{\Delta x}(x, \varphi) = \int_{x-\Delta x/2}^{x+\Delta x/2} dx P(x,t=\pi/\omega)
\end{eqnarray}

\begin{figure}[ht!]
    \centering
    \includegraphics[scale=0.8]{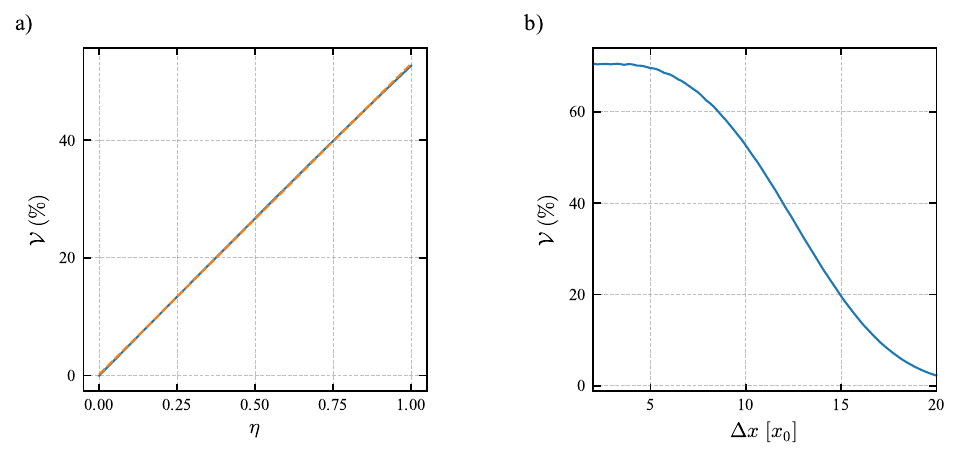}
    \caption{a) Visibility as a function of time-bin coherence parameter $\eta$ (blue line) and linear fit (dashed orange line) giving $\mathcal{V} \approx (52.8\%) \times \eta$.  b) Visibility as a function of detection resolution $\Delta x$. Parameters used are the same as in the main text.}
    \label{fig2_appendix}
\end{figure}

To account for dephasing of the time-bin electron wavefunction we consider an additional environment for the electron, 
\begin{eqnarray}
    \vert \psi_{e^{-}} \rangle = \frac{\vert t_{e}\rangle + e^{i\varphi}\vert t_{\ell}\rangle}{\sqrt{2}}  \rightarrow \vert \psi_{e^{-}}, \mathcal{E} \rangle = \frac{\vert t_{e}\rangle \vert e_{0} \rangle + e^{i\varphi}\vert t_{\ell}\rangle  \vert e_{1}\rangle}{\sqrt{2}}
\end{eqnarray}
where $ \langle e_{0} \vert e_{1} \rangle = \eta $. Fro simplicity we assume $\eta \in \mathbb{R} $. The initial particle + environment state then becomes 
\begin{eqnarray}
    \rho_{S\mathcal{E}, 0} = N^{2} \left( e^{i\frac{\delta p \hat{x}}{\hbar} } \vert e_{0} \rangle + e^{i\varphi} e^{-i\frac{\delta p \hat{x}}{\hbar} } \vert e_{1} \rangle\right) \rho_{th} \left( e^{-i\frac{\delta p \hat{x}}{\hbar} } \langle e_{0} \vert + e^{-i\varphi} e^{i\frac{\delta p \hat{x}}{\hbar} } \langle e_{1} \vert\right)
\end{eqnarray}
Tracing out the environment states leads to the particle's initial state
\begin{eqnarray}
    \langle x \vert \rho_{0} \vert y \rangle = N^{2}\rho_{th}(x,y) \left(  2\cos\left(\frac{\delta p}{\hbar} (x-y)\right)   
 + 2\eta \cos\left( \frac{\delta p}{\hbar}(x+y) - \varphi \right)\right)
\end{eqnarray}
From now on, the calculation mirrors the steps after eq. \eqref{initial_state_appendix}.

Fig. \ref{fig2_appendix}a) shows a plot of the interference visibility as a function of the electron coherence parameter $\eta $, and b) as a function of measurement resolution $\Delta x$.

Finally, we compute the Wigner function at the origin of phase space as a function of time for our initial state \eqref{initial-characteristic-function} specialized to $\varphi = \pi$ and $\delta p = p_{0}$, using \eqref{Wigner-negativity}. Direct calculation yields, 
\begin{eqnarray}
    W(0,0,\tau) = \frac{N^{2}}{2(2\pi)^{2}}\left( I(\hat{A},\vec{b}_{1}) + I(\hat{A},-\vec{b}_{1}) - e^{-\frac{\xi^{2}}{2}}   \left( I(\hat{A},\vec{b}_{2}) + I(\hat{A},-\vec{b}_{2}   \right)\right)
\end{eqnarray}
where we consider $\hbar = 1$, $ \tau = \omega t$, $ N = \left(2 + 2 - e^{-\xi^{2}/2}  \right)^{-1/2}  $, 
\begin{eqnarray}
    I(\hat{A},\vec{b}) = \frac{2\pi}{\sqrt{\det \hat{A}}} \exp\left( \frac{1}{2}\vec{b}^{\rm T} \cdot \hat{A}^{-1}\cdot \vec{b} \right) \ , 
\end{eqnarray}

\begin{eqnarray}
    \vec{b}_{1} &=& \frac{i}{2}\left( \sin \tau, \cos \tau  \right) \\
    \vec{b}_{2} &=& \frac{\xi^{2}}{2}\left( \cos \tau, \sin \tau  \right)
\end{eqnarray}

\begin{eqnarray}
    \hat{A} = \begin{pmatrix}
2\alpha(\tau) + \frac{\xi^{2}}{4} & \gamma(\tau) \\
\gamma(\tau) & 2\beta(\tau) + \frac{\xi^{2}}{2}
\end{pmatrix}
\end{eqnarray}
where $\alpha(\tau), \beta(\tau), \gamma(\tau)$ are defined above.




\end{document}